\documentclass[12pt]{article}

\usepackage{scicite}

\usepackage{times}

\usepackage{amsmath}
\usepackage{amsfonts}
\usepackage{amssymb}
\usepackage{graphicx}

\newenvironment{sciabstract}{%
\begin{quote} \bf}
{\end{quote}}

\title{Fully stabilized multi-TW optical waveform synthesizer for gigawatt soft-x-ray isolated attosecond pulses}

\author
{Bing Xue${}^{1}$, Yuuki Tamaru${}^{1,2}$, Yuxi Fu${}^{1}$, Hua Yuan${}^{3}$, Pengfei Lan${}^{3}$, Oliver D. \\
M\"ucke${}^{4,5}$, Akira Suda${}^{2}$, Katsumi Midorikawa${}^{1}$, and Eiji J. Takahashi${}^{1\ast}$\\
\\
\normalsize{${}^{1}$Attosecond Science Research Team, RIKEN Center for Advanced Photonics,}\\
\normalsize {RIKEN, 2-1 Hirosawa, Wako, Saitama 3510198, Japan}\\
\normalsize{${}^{2}$Department of Physics, Tokyo University of Science, 2641 Yamazaki,}\\
\normalsize{Noda, Chiba 278-8510, Japan}\\
\normalsize{${}^{3}$School of Physics and Wuhan National Laboratory of Optoelectronics,}\\
\normalsize{Huazhong University of Science and Technology, Wuhan, Hubei 430074, China}\\
\normalsize{${}^{4}$Center for Free-Electron Laser Science CFEL, Deutsches Elektronen-Synchrotron DESY,}\\
\normalsize{Notkestra\ss e 85, 22607 Hamburg, Germany}\\
\normalsize{${}^{5}$The Hamburg Centre for Ultrafast Imaging, Luruper Chaussee 149,}\\
\normalsize{22761 Hamburg, Germany}\\
\\
\normalsize{$^\ast$To whom correspondence should be addressed; E-mail:  ejtak@riken.jp}
}

\date{}

\begin{document} 




\maketitle



\begin{sciabstract}
A stable 50 mJ three-channel optical waveform synthesizer is demonstrated and used to reproducibly generate a high-order harmonics supercontinuum in the soft-x-ray region. 
This synthesizer is composed of pump pulses from a 10-Hz-repetition-rate Ti:sapphire pump laser and signal and idler pulses from an infrared two-stage optical parametric amplifier driven by this pump laser. With the full active stabilization of all relative time delays, relative phases, and the carrier-envelope phase, a shot-to-shot stable intense continuum harmonic spectrum is obtained around 60 eV with pulse energy above 0.24 $\mu$J.  
The peak power of the soft-x-ray continuum is evaluated to be beyond 1 GW with a 140 as transform limit duration. 
Furthermore, we found a characteristic delay dependence of the multi-cycle waveform synthesizer and established its control scheme. 
Compared with the one-color case, we experimentally observe an enhancement of the cut-off spectrum intensity by one to two orders of magnitude through the three-color waveform synthesis.
\end{sciabstract}


\section*{Introduction}

Laser technology for producing optical pulses with ultrahigh intensities and ultrashort durations of multiple or few optical cycles has made tremendous progress in recent years, enabling the exploration of various intriguing fundamental mechanisms in strong-field physics and attoscience\cite{mourou_eli_2011}. Pushing the ultrafast and ultra-intense frontier further into the sub-cycle regime has become the next big challenge among the laser science community.
Optical pulses with a sub-cycle duration are an appealing route to sculpt the temporal profile of electric fields $E(t)$, which have a broad variety of prospective applications in strong-field physics, for example, in electron localization\cite{Gulde200,PhysRevA.86.013418}, relativistic laser-plasma interactions\cite{mangles_monoenergetic_2004,dromey_coherent_2012}, plasma wakefield accelerators\cite{geddes_high_quality_2004,faure_laserplasma_2004}, and isolated attosecond pulse (IAP) generation. In particular for strong IAP generation, by overcoming the generated energy bottleneck for attosecond spectroscopy, sub-cycle pulses will definitely initiate a revolution\cite{krausz_attosecond_2014,kraus_ultrafast_2018} in chemical dynamics research\cite{lepine_attosecond_2014,cattaneo_attosecond_2018}. The most promising way to obtain a sub-optical-cycle pulse duration is to synthesize sub-pulses with different colors generated from separate sources\cite{Manzoni}. Currently, the available laser technology is able to realize such synthesizers by using optical parametric (chirped-pulse) amplification [OP(CP)A] or super-continuum generation\cite{Manzoni,rivas_next_2017,7095515,Wirth195}. Experimental efforts have simultaneously achieved a multi-octave-spanning bandwidth, high output intensity, and sub-cycle timing jitter control\cite{huang_high-energy_2011}. However, in these schemes for obtaining such extremely short single/sub-cycle driving pulses, the pulse energy is usually limited to the 10-microjoule class at kHz repetition rates\cite{huang_high-energy_2011,haessler_enhanced_2015}. Consequently, the generated IAP is usually limited to the sub-nanojoule level, which provides insufficient photon flux required for true attosecond pump/attosecond probe experiments. While continuous effort to realize the high energy IAP has been reported using advanced schemes such as the formation of averaged IAPs\cite{takahashi_attosecond_2013,PhysRevLett.104.233901} from a two-color laser field, there is no demonstration for ideal/perfect IAP energy-scaling result so far.

Theoretically, synthesized sub-cycle driver pulses with a ``perfect waveform'' were proposed for optimized high-order harmonics generation (HHG)\cite{PhysRevLett.102.063003} and IAP production\cite{Zeng_PRL_2007,Lan_PRA_2007}, and the enhancement of the HHG by using such synthesized waveforms was experimentally and theoretically studied\cite{PhysRevX.4.021028,jin_waveforms_2014,jin_route_2014}. On the basis of the above results, synthesized few/multi-cycle driving pulses are expected to enable scaling up of the energy and efficiency of IAP generation\cite{PhysRevX.4.021028}. To obtain IAPs with pulse energies reaching the microjoule level, a TW-class waveform synthesizer is necessary, which requires the use of high-energy femtosecond lasers, that are all operated at low repetition rates (e.g., 10 Hz). The technical challenges of developing such a large synthesizer system and in particular to stabilize the laser waveform are tremendous. This is due to the low sampling rates required for actively eliminating relative timing and phase jitters as well as stabilizing carrier-envelope phases (CEPs) for each channel in the synthesizer.

In this work, we demonstrate a multi-TW three-channel parallel parametric waveform synthesizer with complete stabilization of all synthesis parameters for the first time, consisting of pulses with 800-, 1350-, and 2050-nm wavelengths. Most importantly, to realize full stabilization of the synthesizer, we have newly developed a CEP-stabilized high-energy Ti:sapphire chirped-pulse amplifier\cite{Takahashi:15} operating at 10 Hz and a phase-and-timing stable frontend. By innovatively introducing multi-cycle pulses synthesis strategy, and combining the waveform synthesizer and the loose-focusing method for HHG, a \textit{shot-to-shot-reproducible} high-energy ($\sim$ 0.24 $\mu$J) harmonic continuum in the soft-x-ray region is achieved. The peak power of the soft-x-ray continuum is evaluated to be beyond 1 GW with a 140 as transform limit duration.
To our knowledge, our obtained photon flux of the super-continuum in the soft-x-ray region is the highest value ever reported. By using a custom-tailor electric field for optimizing HHG intensity, we have established a perfect IAP energy-scaling for the first time. On the basis of the achieved performance, we foresee attosecond spectroscopy applications using both attosecond pump and probe pulses in the near future.

\section*{Results}

\label{sec:synthesizer}

Our synthesized three-channel waveforms consist of a multi-cycle Ti:sapphire laser pulse and two multi-cycle infrared OPA signal and idler pulses. 
The two-stage OPA is pumped by part of the energy of the Ti:sapphire laser. 
Figure S1 shows the system scheme of the high-energy three-channel waveform synthesizer; note the pump pulse with a wavelength centered at 800 nm, the signal pulse centered at 1350 nm, and the idler pulse centered at 2050 nm. 
More details are provided in the Methods Section. 
The durations of the pump, signal, and idler pulses at the synthesis output are measured separately by different methods. We use spectral phase interferometry for direct electric-field reconstruction for the pump pulse and a frequency-resolved optical gating for signal and idler pulses. 
Detailed results are shown in Figs. S2(a), (b), and (c). 
The pump (30 fs FWHM), signal (44 fs), and idler (88 fs) pulses are considered to be well compressed owing to their durations approaching the corresponding Fourier transform limits (FTLs). 
The spectral combination of the three pulses is shown in Fig. S2(d).

\subsection*{Synthesizer stabilization}

To custom-tailor the electrical fields using three-channel synthesis, we need to stabilize both the CEPs ($\varphi_{\rm CEP}$) and the relative timings among the three channels as mentioned in the introduction. The combined three-channel electrical field $E_{\rm 3c}$ can be expressed by
\begin{equation}
 E_{\rm 3C} = \sum_{d=p,s,i}E_d\exp\left[-2\ln 2\left(\frac{t+\delta t_d}{\tau_{d}}\right)^2\right]\cos\left(\omega_d\left(t+\delta t_d\right)+\varphi_d\right)\\
\label{eq:equation1}
\end{equation}
\noindent where the subscript $d$ ($p$, $s$, $i$) denotes the pump, signal, and idler pulses, respectively. 
$E_d$ is the electric field amplitude; $\tau_{d}$ the full width at half maximum (FWHM) pulse duration; $\omega_d$ the carrier frequency; $\delta t_d$ is defined as the time delay between each pulse and the reference (signal) pulse, therefore $\delta t_s$ is set to zero; as the pump pulse is at the fundamental frequency, the CEP of the pump pulse $\varphi_p$ is equal to that of the Ti:sapphire laser ($\varphi_{\rm CEP}$). The two-stage IR OPA is seeded by a white-light continuum generated in a sapphire plate in the first stage OPA-1; the CEP of the signal pulse is determined by the pump pulse and preserved in the amplified signal pulse. 
Therefore, the signal pulse CEP can be defined as $\varphi_s$ = $\varphi_{\rm CEP}$ + $\varphi_A$, where $\varphi_A$ is an active value, which is determined by the delay jitter introduced by optical-path-length and pulse-energy fluctuations combined with amplitude-to-phase noise conversion. 
The idler pulse $\varphi_{i}$ has a constant value due to self-CEP-stabilization in the difference-frequency generation process in the OPA\cite{PhysRevLett.88.133901,Cerullo_LPR11}. The typical single-shot CEP noise ($\varphi_{\rm CEP}$) of our 10-Hz Ti:sapphire laser, measured by single-shot f-to-2f interferometry (which will also be used later for CEP tagging in Fig. 6), has been stabilized\cite{Takahashi:15} to 480 mrad root-mean-square (rms), as shown in Fig. 1(a).

\begin{center}
\includegraphics[width=0.8\linewidth]{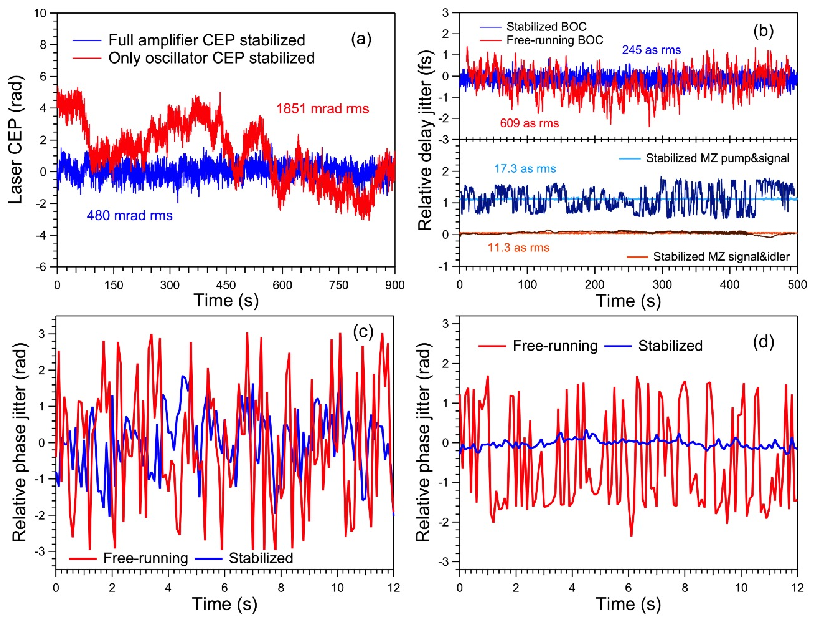}\\
\end{center}
\noindent {\bf Fig. 1. Stabilized parameters of the synthesizer.} (a) 10-Hz Ti:sapphire laser CEP values measured by a laboratory-built f-to-2f interferometer
when the complete amplifier system is CEP-stabilized (blue curve, improved to 480 mrad rms) compared with the case that only the amplifier's seed oscillator is stabilized (red curve). (b) Lower panel: delay jitter performance of Mach--Zehnder interferometer stabilization between pump and signal path (improved to 17.3 as rms), signal and idler path (improved to 11.3 as rms); upper panel: delay jitter performance with balanced optical cross-correlator between pump and signal path (improved from 609 as rms in the free-running case to 245 as rms when stabilized). Relative phase jitter performance with (blue curve) and without (red) stabilization between pump and signal pulses (c, improved to 816 mrad rms) and signal and idler pulss (d, improved to 106 mrad rms).

To realize a shot-to-shot-stable synthesized electric field, the field amplitudes, CEPs, and temporal delays must all be stabilized. 
In Table S1, we summarize all the stabilized parameters in this three-channel synthesizer system. The signal pulse is defined as the reference pulse, and all the jitter parameters are listed as the differences with respect to the signal reference pulse.
For stabilization of the delay jitter, Mach--Zehnder (MZ) interferometers and balanced optical cross-correlators (BOCs)\cite{Manzoni} are implemented (see Methods Section). MZ interferometers are a helpful tool for suppressing the delay jitter between all optical channels due to mechanical vibrations of mirrors and air turbulence, as shown in the lower panel of Fig. 1(b). Furthermore, pulse energy fluctuations of the Ti:sapphire laser are another source of nonnegligible delay jitter, as delay jitter is introduced during the white-light generation process in OPA-1 by the nonlinear Kerr effect\cite{1250454}. For this type of delay jitter, an in-loop BOC is introduced to measure and feedback-stabilize the delay between the pump and signal pulses. With the closed feedback loop in the delay line of the seed pulse before OPA-2, operating at a frequency of 10 Hz, this delay jitter is precisely suppressed to $\sim$245 as rms during 500 s operating time, as shown in the upper panel of Fig. 1(b). The delay jitter between signal and idler pulses is also measured by another out-of-loop BOC setup, which shows small fluctuations; by taking into account the resolution of the balanced signal, it unambiguously confirms the achieved attosecond-precision relative timing stability\cite{Manzoni}. The relative phase jitter is also an important quantity for evaluating the stability of synthesis. We show the measured relative phase jitters between pump and signal pulses in Fig. 1(c) and between signal and idler pulses in Fig. 1(d). They are considered to be passively stable when the laser CEP and all the delay jitters are actively stabilized. A more detailed discussion of the relationship between delay jitter and relative phase jitter will be provided later. 
The CEP jitter value shown in Table S1 for the signal pulse is estimated by adding the pump pulse CEP jitter and the relative phase jitter between the pump and signal corresponding to the measured delay jitter (BOC).

\subsection*{Stable high-order harmonic super-continuum generation}

To prove the effectiveness of our synthesizer stabilization scheme, we demonstrate the performance of the synthesizer in the application of HHG. A loose focusing geometry with a long medium length\cite{Takahashi:03} is employed to scale up the HHG output energy. 
The sub-pulses of the synthesized waveforms are focused by using two separate long-focal-length lenses (4.5 m for pump pulse, 3.5 m for signal and idler pulses) into an 8-cm-long argon gas cell. 
The focused intensity in the three-color case is measured to be $1\times10^{14}$  W/cm$^{2}$ with input pulse energies of 20.3, 4.3, and 1.6 mJ for the pump, signal, and idler, respectively. 
The generated HH is then spectrally dispersed by an aberration-corrected concave grating and detected by a combination of a microchannel plate (MCP) and a fluorescent screen with a CCD camera. 
The HH yield is carefully optimized by adjusting the focusing point while varying the gas pressure until the phase-matching condition is fulfilled. 
The continuum spectrum intensity of three-color HHG is finally optimized at a gas pressure of 2.8 Torr, which is in good agreement with a calculation based on phase-matching theory\cite{PhysRevLett.82.1668} under the current experimental conditions. Figure 2(a) shows single-shot HH spectra when one-, two-, and three-color driver waveforms are used. In the one-color case, clear discrete harmonics are generated with the highest intensity peak at the 25th order around 39 eV. 
Also, the harmonics end at the 35th order with a photon energy of approximately 54 eV. With two-color (pump and signal) driving, a super-continuum spectrum appears in the cut-off region (above 50 eV). However, a strong interference pattern is still visible in the spectrum, which indicates the existence of satellite pulses around the main HH pulse. 
When the third idler pulse is also  injected, a smooth continuum harmonic spectrum is observed and the HH cut-off region is further extended to 70 eV. 
The modulation of the spectrum disappears because the largest electric field (pump) extrema on both sides of the central peak are clearly suppressed in amplitude by the signal and idler pulses. 
Taking our previous experimental results into account\cite{doi:10.1063/1.1637949}, the measured continuum soft-x-ray (50--70 eV) pulse energy is evaluated to exceed 0.24 $\mu$J with a conversion efficiency of $10^{-5}$ at the generation point. Compared with the one-color case, we experimentally observed the enhancement of the cut-off spectrum intensity by one to two orders of magnitude through the three-color waveform synthesis.
The bandwidth of this generated continuum spectrum supports a 140 as pulse. This FTL pulse is shown in the inset of Fig. 2(a). Actually, we can expect to obtain a nearly FTL pulse because the attosecond chirp is negligible in the cut-off region.

\begin{center}
\includegraphics[width=0.9\linewidth]{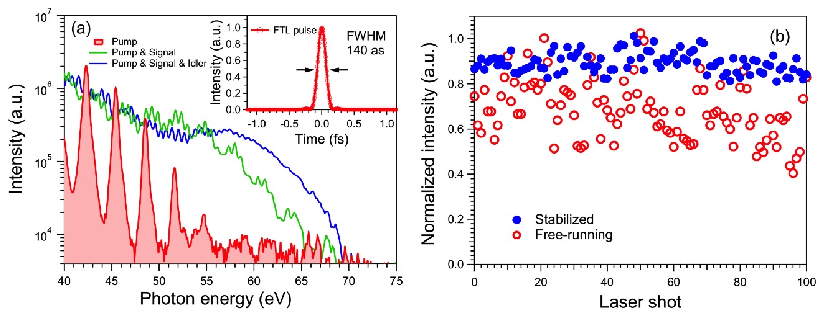}\\
\end{center}
\noindent {\bf Fig. 2. High-order harmonics supercontinuum generation.} (a) Measured single-shot high-order harmonics spectra for one- (red curve, pump only), two- (green, pump and signal), and three-color (blue, all) driver waveforms. The inset shows the Fourier-transform-limited pulse from the continuum spectrum (50--70 eV) in the cut-off region. (b) Stability of the intensity integrated over the continuum spectrum (50--70 eV) with and without stabilization of the synthesizer; when stabilized, the rms noise of the intensity of the continuum spectrum is reduced from 15.1\% rms (red circles) to 4.1\% rms (blue dots).

When the complete stabilization scheme of the synthesizer is working effectively, the stability of the generated high-order harmonics is significantly improved. 
In Fig. 2(b) we show a comparison of the performance of the intensity stability under free-running and fully stabilized (CEP, MZ, and BOC) conditions by integration of the continuum spectrum (50--70 eV) intensities for each individual laser shot. 
The intensity of the cut-off region under the fully stabilized condition exhibits good stability, as shown by the blue points, whereas it fluctuates between 0.4 and 1 under free-running condition (red circles).
In the free-running case, the generated HH spectrum intensity randomly changes because of the nonreproducible complex electric field due to the unstabilized CEPs and delay jitters. 
With the CEP, MZ interferometer, and BOC stabilization active, we can effectively improve the stability of the synthesized electric-field waveform. 
As a result, the relative intensity variation of the harmonics is clearly improved, while the single-shot spectrum intensity fluctuation is decreased from 15.1\% rms to 4.1\% rms. This is remarkable stability considering the high-order nonlinearity of the generation mechanism.

\subsection*{Pulse delay time dependences}

By inserting the measured values of the synthesis parameters into Eq. (\ref{eq:equation1}), a simulated synthesized electric-field waveform $E(t)$ and the related instantaneous intensity profile $E^2(t)$  (note this is not the textbook definition of intensity, which is a cycle-averaged quantity)\cite{Wirth195} for the three-color case can be obtained, as depicted in Fig. S3. The CEPs are all assumed to initially be zero.  As an immediate result under this condition, the central peak of the three-color synthesized electric field lasts only 0.32 cycles (in amplitude FWHM) of the carrier frequency (800 nm). The peak intensity ratio between the main central peak and the nearest-neighbor peaks is increased from 1.02 in the one-color case to 2.1 in the three-color case. Note that this intensity ratio is comparable to that of a 2.5-fs pulse in the 800 nm one-color case (i.e., pulse duration below one optical cycle). Thus, the neighbor peaks are effectively suppressed in contrast by the waveform synthesis. Owing to this effective neighbor-peak suppression, an isolated pulse with no side peaks can be generated in the three-color case as shown in the inset of Fig. 2(a), which has also been experimentally confirmed by the disappearance of the fringes in the HH super-continuum spectrum.

As mentioned in the introduction, it was theoretically predicted that a broadband and enhanced HHG can be obtained by a synthesized ``perfect driver waveform''\cite{PhysRevLett.102.063003,jin_route_2014,jin_waveforms_2014}. According to Eq. (\ref{eq:equation1}), the synthesis parameters are the field amplitudes, phases, and delay times of all channels. By gradually changing the set point of each stabilized parameter, we can tailor a customized shot-to-shot reproducible waveform of the synthesized electric field. This experimental capability now opens the door to finding the best waveform for optimizing certain aspects of HHG\cite{jin_waveforms_2014}, such as, the ideal waveform for most efficient and highest-energy HHG. In particular, in our multi-cycle waveform synthesizer, when the delay change $\delta t_d$ in Eq. (\ref{eq:equation1}) is much smaller than the pulse duration $\tau_{d}$, the synthesized electric field can be rewritten as
\begin{equation}
 E_{3C} \cong \sum_{d=p,s,i}E_d\exp\left[-2\ln 2\left(\frac{t}{\tau_{d}}\right)^2\right]\cos\left(\omega_d\left(t+\delta t_d\right)+\varphi_d\right)\,.\\
\label{eq:equation2}
\end{equation}
In Eq. (\ref{eq:equation2}), changing the delay of a sub-pulse only affects its phase part in the argument of the cosine function. Therefore, while all other related parameters are stabilized in the synthesizer, a delay change ($\delta t_d$) is in fact equivalent to a CEP variation ($\varphi_d$) for each sub-pulse. 
This feature has very important consequences as it strongly benefits the experimental realization. The absolute-value control of the CEPs is recognized to be experimentally difficult and complicated in a real experimental setup, particularly for few- or sub-cycle waveform synthesizers, in which all CEPs and relative timings must be \emph{simultaneously} stabilized\cite{Manzoni,7095515}. 
In contrast, in a multi-cycle synthesizer, the challenging prerequisite of simultaneous control of all CEPs and relative timings is, via Eq. (\ref{eq:equation2}), reduced to control of the delays only, which is much easier to achieve experimentally. 
We emphasize once again that this feature requires delay changes much smaller than the pulse duration, which is only satisfied in a multi-cycle synthesizer. 

\begin{center}
\includegraphics[width=0.9\linewidth]{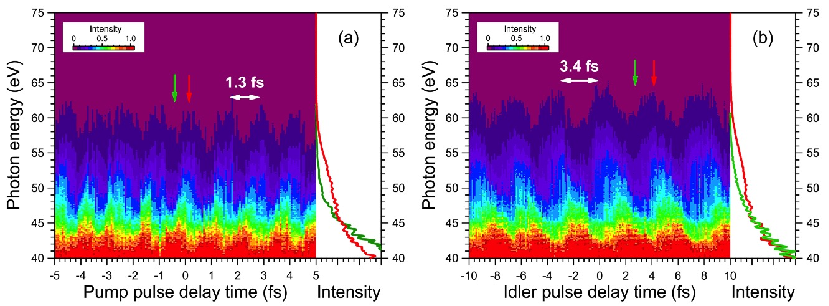}\\
\end{center}
\noindent {\bf Fig. 3. Delay time dependence of the generated single-shot harmonics continuum spectrum.} (a) Experimental result when only changing the delay of the pump pulse; (b) the same for the idler pulse. The insets on the right-hand sides show the corresponding single-shot harmonics continuum spectra for the delay times marked by the red and green arrows.

In Fig. 3, we show the experimentally obtained delay time dependence of the generated high-order harmonics spectrum by changing either the pump (Fig. 3(a)) or idler pulse delay (Fig. 3(b)). 
The pump pulse delay dependence is measured by stabilizing the laser CEP and the MZ interferometer (signal and idler), and by varying the delay of the pump pulse in steps of 50 as. 
On the other hand, the idler pulse delay dependence is measured by stabilizing the CEP, MZ interferometer (pump and signal), pump pulse timing using the BOC (pump and signal), and scanning the idler pulse delay in steps of 100 as.
Both delay-scan maps in Fig. 3 show that the delay of each sub-pulse strongly affects the HH spectrum, especially the cut-off continuum spectrum. In particular, at certain delays, not only is the spectrum extended, but also the energy is enhanced.
From the cosine function in Eq. (\ref{eq:equation2}), one can intuitively expect that the time periodicity for a delay change of each sub-pulse will be one optical cycle ($T_d=1/(2\pi\omega_d)$). However, while the delays of pump and idler pulses change separately, the generated spectra both exhibit periodic changes in half a cycle (1.3 fs for pump, 3.4 fs for idler). 
Note, that the delay-scan map for the idler shows less noise and better data quality. 
This is because the idler CEP is a constant value owing to the passive self-stabilization of the OPA idler\cite{PhysRevLett.88.133901,Cerullo_LPR11}. 
In the other pump pulse case, BOC stabilization is not active while scanning the pump delay; thus, we see the effects of imperfect stabilization in the HHG spectra in Fig. 3(a). 
Moreover, these results show excellent repeatability between daily experiments. 
This was initially a concern for us because the value of the laser CEP changes on a daily basis even though it is well stabilized. 
Strictly speaking, when the laser CEP value changes, the synthesized electric waveform also changes. Good repeatability from day to day suggests that some underlying rules are effective, details of which will be discussed later.

\begin{center}
\includegraphics[width=0.9\linewidth]{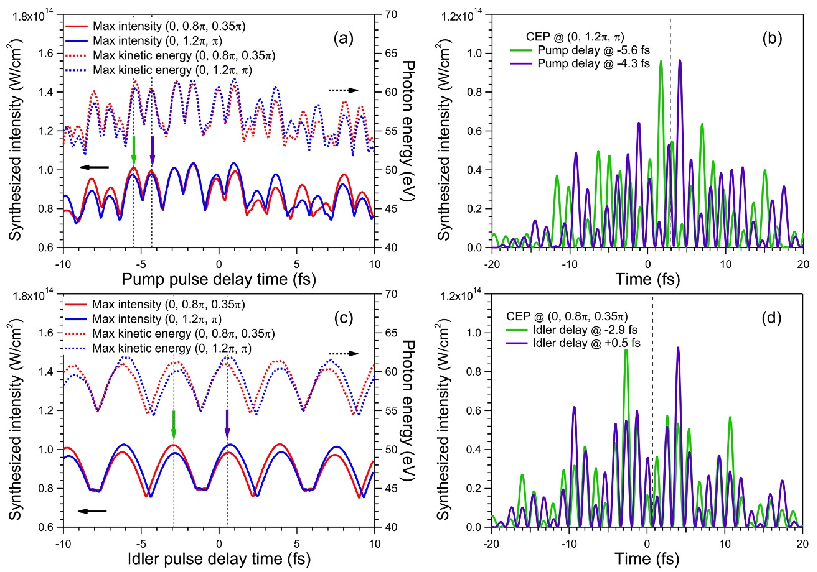}\\
\end{center}
\noindent {\bf Fig. 4. Simulation of synthesized waveform.} Simulated maximum instantaneous intensity profile $E^2(t)$ (solid curves) of the synthesized waveform and maximum kinetic energy profile (dotted) of the HH spectrum in the cut-off region when (a) only the pump pulse delay is varied, (c) only the idler pulse delay is varied. (b) and (d) show the corresponding instantaneous intensity profiles for two different pump pulse delays corresponding to the maximum-intensity peaks indicated by red and blue arrows in (a) and (c), respectively.

The extension of the HHG continuum spectrum is related to the maximum intensity of the synthesized electric field. 
To understand the half-cycle periodicity better, the synthesized instantaneous intensity is simulated by varying the delay parameters in Eq. (2). 
By substituting the experimentally measured parameter values into Eq. (2), the unknown absolute values of the CEPs are considered to be variable parameters. 
When the CEPs are set to (0, 0, 0), referring to the CEPs of the pump, signal, and idler, respectively, the maximum instantaneous intensity appears in the center of the synthesized waveform when all the delay times are set to zero, as shown in Fig. S3(b). 
This maximum-intensity peak shows one-cycle periodicity when the delay of the pump or idler changes. However, this periodic behavior collapses if a change in the initial CEPs occurs. 
The maximum synthesized intensity is finally found to be half-cycle periodic by assigning specific CEP values to all sub-pulses in the simulation, as shown in Figs. 4(a) and (c). 
In contrast to the (0, 0, 0) case, a maximum in intensity alternately appears at two different positions on a timescale over which the delay of the pump or idler changes, as shown in Figs. 4(b) and (d). A conjugate line can be found between the two synthesized waveforms (vertical dotted lines in Figs. 4(b) and (d)). 
The maximum peak intensity of each individual waveform (blue or green) exhibits a one-cycle periodic feature. 
However, when scanning the delay in Figs. 4(a) and (c), the maximum peak appears twice per cycle (i.e, the feature becomes half-cycle periodic) as the HHG contributions from the blue and green waveforms alternate.
In addition, two sets of CEP combinations are found, for which the theoretical results are in good agreement with the experimental ones, which are (0, 0.8$\pi$, 0.35$\pi$) and (0, 1.2$\pi$, $\pi$). 
Note that the sum of the phases of these two signal pulse is exactly 2$\pi$. Moreover, the same results can be obtained when adding or subtracting 2$\pi$ to the 0.8$\pi$ and 1.2$\pi$ phases (i.e., for -0.8$\pi$, -1.2$\pi$, 2.8$\pi$, 3.2$\pi$, etc.). There should be no other suitable combination of CEPs in the 0 to 2$\pi$ range. These results provide a way of back-tracing the exact CEP values of the sub-pulses, which otherwise cannot be directly measured in our experiment. By using the identified CEP values, we can then calculate the maximum kinetic energy of the continuum spectrum as a function of the pump or idler delay time on the basis of the classical three-step model of HHG\cite{CorkumPRL1993}. The results are shown as dotted curves in Figs. 4(a) and (c). The shapes of these curves approximately agree with the maximum-intensity curves of the simulated synthesized waveform. 
A change in the laser CEP clearly affects the synthesized electric field. However, by appropriately changing the initial delays of the pump and idler pulses, similar delay variation characteristics can be recovered and traced back to the initial CEP of the laser. This result has great significance for our experiments: while the laser CEP is stabilized to different absolute values each day, the same synthesized waveform can always be obtained by appropriately changing the delays of the pump and idler pulses. This conclusion has been experimentally confirmed by observing the same delay dependence in results on different days.

We also perform theoretical calculations of HHG spectra in this work. The numerical simulations are carried out by considering the single-atom response induced by the three-color synthesized electric field, and the collective response of the macroscopic gas target to the laser driving field and high-harmonic fields\cite{PhysRevA.82.053413,PhysRevA.79.043413}. The calculations are based on the strong-field approximation (SFA) model\cite{SFA1994}, and the collective response is calculated by solving the propagation equation\cite{PhysRevA.82.053413,PhysRevA.79.043413}. The parameters used in our numerical calculations are directly taken from the experimental measurements and results, and the absolute CEP values of the sub-pulses are (0, 0.8$\pi$, 0.35$\pi$) and (0, 1.2$\pi$, $\pi$), as obtained from the previous electric field simulation results shown in Fig. 4. When changing the delay of the pump pulse, as shown in Figs. 5(a) and (b), the maximum cut-off spectrum extension appears with a periodicity of 1.3 fs. This delay time dependence exactly agrees with our expectation. In the calculation result for changing the delay of the idler, the shown features are in even better agreement with the experiment, as shown in Figs. 5(c) and (d).

\begin{center}
\includegraphics[width=0.9\linewidth]{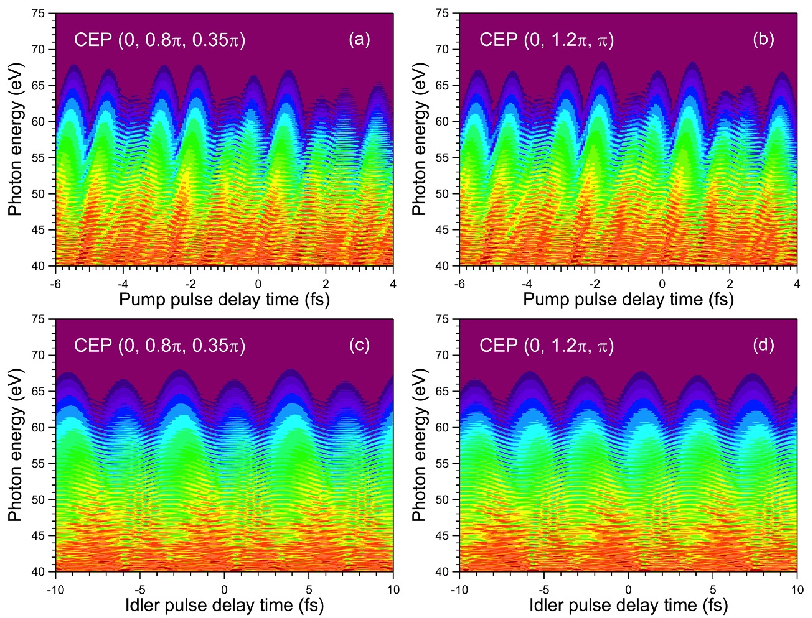}\\
\end{center}
\noindent {\bf Fig. 5. Theoretical calculation results of HHG spectrum profile.} By using the synthesized electric-field waveforms, when only changing the pump pulse delay in (a), (b); the same for only changing the idler pulse delay in (c), (d). The absolute values of CEPs in (a) and (c) are (0, 0.8$\pi$, 0.35$\pi$) and those in (b) and (d) are (0, 1.2$\pi$, $\pi$).

\subsection*{Laser CEP dependence of HHG spectra}

As mentioned in the previous section, the CEPs of pump, signal, and idler pulses affect the synthesized electric-field waveform. Moreover, the idler CEP has a constant value, and the pump and signal CEPs depend on the laser CEP. Therefore, in the case of three-color synthesis, the laser CEP dependence of the HHG spectrum becomes more complicated than that in the one- or two-color cases. It also strongly depends on the stabilization performance of the synthesizer. More specifically, according to Eq. (2), the CEPs can be modified by changing the delays of each sub-pulse. Therefore, the laser CEP is not an independent parameter for the HHG continuum spectrum; thus, a simple, unique laser CEP dependence does not exist. However, we can still expect an interpretable laser CEP dependence under certain conditions. In Fig. 6(d), the maximum intensity of the synthesized electric field is simulated as a function of the laser CEP. To consider three typical stabilized conditions of the synthesizer, we define the initial CEPs of each channel as (0, 0, 0), (0, 0.8$\pi$, 0.35$\pi$), and (0, 1.2$\pi$, $\pi$). Under these three conditions, the laser CEP dependence exhibits three different patterns. When the initial CEPs are (0, 0, 0), the harmonic continuum cut-off extension occurs only once during a 2$\pi$ change in the laser CEP. In contrast, for the other two conditions, the cut-off extension occurs twice. As we already know from our previous discussion, the maximum intensity of the synthesized electric field is related to the continuum spectrum extension; these experimental results also show good agreement with these three different intensity patterns. In Figs. 6(a), (b), and (c), the laser CEP dependence of the HHG spectra corresponds to the initial CEPs of (0, 0, 0), (0, 0.8$\pi$, 0.35$\pi$), and (0, 1.2$\pi$, $\pi$), respectively. These results are measured using CEP tagging, which was frequently used in previous works\cite{PhysRevA.80.061402}. 

In Figs. 6(b) and (c), it is clear that the intensity of the continuum spectrum (between 45 and 65 eV) is much stronger than that in Fig. 6(a). This is another important feature guiding us when performing the experiment. During the optimization of the synthesizer by varying the delays of the pump and idler pulses, the only observable is the generated harmonic continuum spectrum. Optimizing the parameters to obtain a higher intensity of the continuum spectrum will automatically lead us to set the CEP values to (0, 0.8$\pi$, 0.35$\pi$) or (0, 1.2$\pi$, $\pi$). Therefore, when conducting the experiment on different days, we can always find reproducible delay dependences for the synthesizer, regardless of the absolute value to which the laser CEP is stabilized on that day.

\begin{center}
\includegraphics[width=0.9\linewidth]{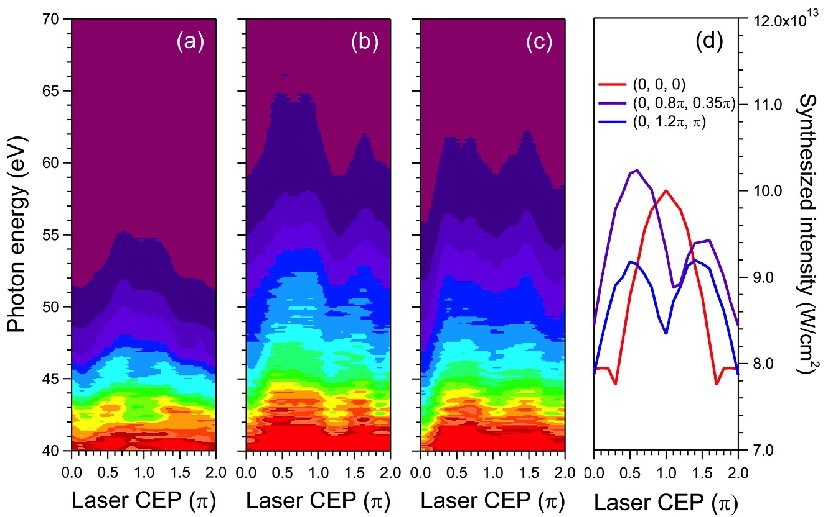}\\
\end{center}
\noindent {\bf Fig. 6. Laser CEP dependences.} Generated high-order harmonics spectrum measured using CEP tagging  for the back-traced initial CEPs of (0, 0, 0) in (a), (0, 0.8$\pi$, 0.35$\pi$) in (b), and (0, 1.2$\pi$, $\pi$) in (c). (d) Simulated maximum intensity of the synthesized electric field versus laser CEP.

\section*{Discussion}

How to obtain a stable and repeatable synthesized electric-field waveform is the key point in using our synthesizer in further applications. In this work, we have experimentally proven the active stabilization of all delay jitters and the laser CEP. As discussed, in our multi-cycle synthesizer, the complexity of the required stabilization scheme is greatly reduced (only precise control of all sub-pulse delays is required) and the scheme is facilitated compared with a few- or sub-cycle synthesizer. As a result, we obtained a stable high-photon-flux HHG continuum in the soft-x-ray region. Comparing the effects of stabilization of the pump and idler pulses, the pump pulse is more difficult to stabilize and less stable than the idler pulse. It may be concluded that a shorter wavelength requires an even higher accuracy in the delay control to achieve the same phase change. From this argument, it is expected that increasing the main driver pulse (pump) wavelength will allow greater stability to be achieved. At the same time, a longer driver wavelength supports a higher cut-off photon energy in HHG and the generation of shorter attosecond pulses\cite{ChenE2361,Takahashi_waterwindow_2008}.

Finally, we demonstrated a 50-mJ-level intense three-channel optical waveform synthesizer with complete stabilization of all synthesis parameters consisting of multi-cycle pulses at 800-, 1350-, and 2050-nm wavelengths. 
The three-color synthesizer was experimentally confirmed to be capable of generating a shot-to-shot-reproducible high-flux broadband continuum spectrum in the soft-x-ray region at 50--70 eV in argon gas. 
The energy of the generated HH continuum spectrum was measured to be above 0.24 $\mu$J, and the continuum spectrum supports a 140-as pulse duration. 
To our knowledge, these soft-x-ray pulse parameters represent a record performance for what has so far been achieved in experiments. 
The intensity stability of the HHG spectrum was improved by the stabilization of the synthesizer to 4.1$\%$ rms for single-shot HHG output, which is remarkable stability considering the high-order nonlinearity of the generation mechanism. 
Furthermore, the delay dependences of the pump and idler pulses on the HHG spectrum was experimentally studied and confirmed by theoretical modeling. 
The control of all CEPs is facilitated in our multi-cycle waveform synthesis scheme. 
We obtained an enhanced harmonic continuum spectrum output by precision delay control that, in the multi-cycle regime, is equivalent to precision CEP control. These observations also provide strong evidence for the fully controlled sub-cycle electric-field waveforms generated by our synthesizer. This work finally paves the way for further attoscience applications and for realizing attosecond pump/attosecond probe spectroscopy in the soft-x-ray region. 
Our three-color multi-cycle synthesis scheme also allows intense attosecond pulses to be generated in the water-window region when using intense longer-wavelength IR driver pulses as the main pulse in the multi-color synthesis\cite{ChenE2361,Takahashi_waterwindow_2008}.

\section*{Methods}
%

\subsection*{Optical waveform synthesizer setup}
For the three-channel optical waveform synthesizer (see Fig. S1), the driver source is provided by a TW-class Ti:sapphire multipass power amplifier (MPA) pumped by two 10-Hz Nd:YAG lasers. 
The CEP-stable seed pulse of this MPA is provided by a commercial CEP-stable Ti:sapphire CPA system with 1 kHz repetition rate (7 mJ, 25 fs). 
The beam pointing of the 1-kHz seed pulse is actively locked by a beam-pointing stabilizer for long-term stable operation.  
After the grating pair compressor, a 170-mJ (energy fluctuation 1.4\% rms) pulse energy is used for the synthesizer system, which is only part of the total energy (800 mJ) from the MPA.
The pump pulse channel of the synthesizer is directly obtained from part of the energy after the compressor without a spectrum change, and precisely compressed with a 12-mm-thick fused silica plate.
All reflection mirrors for the pump pulse before the down-collimation for the final waveform synthesis are chosen to have 3-inch diameter to avoid laser damage. A motor-controlled stage and a piezo-actuated delay stage are installed in the pump pulse path to adjust and fine-tune the delay of the pump pulse. The output energy of the pump pulse in the synthesis is up to 44 mJ. The signal and idler pulses are generated by a two-stage infrared OPA. 
The first-stage OPA (OPA-1) is a commercial white-light-seeded OPA system, which is pumped by a 2-mJ fundamental pulse after spatial shaping. 
OPA-1 provides a weak infrared seed pulse (200 $\mu$J) for the second-stage OPA (OPA-2). Before seeding OPA-2, the seed pulse is optically delayed by a piezo-actuated delay stage and its beam size is expanded. OPA-2 is a laboratory-built collinear OPA, which is pumped by a 24-mJ 800-nm down-collimated laser pulse.
 From OPA-2, intense signal (6.1 mJ, energy fluctuation 2.1\% rms) and idler (4.3 mJ, energy fluctuation 2.6\% rms) pulses can be obtained. 
 The nonlinear medium for OPA-2 is a 2-mm-thick type-II BBO crystal. 
 Owing to the type-II phase-matching condition, the signal and idler pulses have orthogonal polarizations. 
 Therefore, we separate the signal and idler pulses by using a dichroic mirror and change the polarization of the signal pulse to be the same as that of the pump and idler pulses with a half-wave plate. 
 Then, the signal and idler pulses are recombined by another dichroic mirror. 
 In the idler pulse path, a piezo-actuated delay stage is used to fine-tune the delay between the signal and idler pulses. 
 Finally, after recombining the signal and idler pulses with the pump pulse using a dichroic beam splitter, all three pulses are collinearly combined. The total output energy reaches to 54 mJ. If all phases of the pulses are set to be zero, the synthesized peak power is estimated to reach 2.6 TW. Then, part of the energy is sent to the HHG experiment.

\subsection*{Mach--Zehnder interferometer stabilization}
To stabilize the optical path length variations among all delay lines, we inject two continuous-wave (CW) laser beams into the synthesizer setup, which uses two MZ interferometers. One CW laser (He-Ne laser at 633 nm) is injected into the signal-pump MZ interferometer, and the other (He-Ne laser at 543 nm) is injected at the first dichroic mirror, where the signal and idler pulses are separated. By using photodiodes after a slit filter, we can monitor the intensity fluctuations of the spatial interference fringes on the recombined beams resulting from optical path length variations. Then, by feedback-controlling piezo-actuated delay stages, which are installed in the pump and idler paths, we successfully suppressed the delay jitters (to less than 20 as rms) in both MZ interferometers. Detailed experimental results are shown in Fig. 1(b). These jitter values in the optical paths are sufficiently small to stabilize the synthesized waveform.

\subsection*{Balanced optical cross-correlators}
In this work, one in-loop BOC and one out-of-loop BOC are introduced separately for feedback stabilization and monitoring the actually achieved delay jitters between the pump and signal, and signal and idler pulses, respectively.  The reference beams used for the BOC measurements are collected from leaked beams after the dichroic mirrors, where the beams are recombined. The detailed implementation of the BOC setups is similar to that described in Ref. 10. In particular, the delay jitter between signal and idler pulses is mainly related to optical path length variations. Therefore, the delay fluctuations in the monitored BOC signals are small and can easily be stabilized by MZ interferometers. Experimentally, with the MZ interferometer feedback active, the delay jitter between signal and idler pulses, as measured by the monitoring BOC, is suppressed to 137 as rms.

\subsection*{Relative phase jitter measurements}
The measurement of phase jitter between pump and signal pulses is achieved by spectral broadening of the signal pulse and by monitoring the variations of the interference fringes between the broadened signal and pump spectra with a spectrometer. The relative phase jitter between signal and idler pulses is measured using the interference fringes appearing in the spectral overlap region of their second harmonics. As stabilized results, achieved when the previously discussed stabilizers (CEP, delay jitters) are active, the relative phase jitters are suppressed to 816 and 106 mrad rms for pump--signal and signal--idler pulses, respectively. The phase jitter between the pump and signal pulses is larger than expected. We speculate that this rather high value is due to a latent measurement error caused by the pulse energy fluctuations during the white-light generation from the signal pulse.


\section*{Supplementary Materials}
Fig. S1. System scheme of the three-channel optical waveform synthesizer.\\
Fig. S2. Sub-pulses composing the synthesized waveforms.\\
Fig. S3. Simulation of the synthesized electric-field waveform.\\
Table S1. Stabilized parameters of the  three-channel optical waveform synthesizer.\\

\bibliography{reference}
\bibliographystyle{ScienceAdvances}

\noindent \textbf{Acknowledgements:} 
%
\\
\noindent \textbf{Funding:} 

This work was supported in part by the Ministry of Education, Culture, Sports, Science and Technology of Japan (MEXT)  through Grant-in-Aid under Grant 16K13704 and Grant 17H01067, in part by the FY 2018 Presidents Discretionary Funds of RIKEN, and in part by the Matsuo Foundation.
K. M. acknowledges support by MEXT Quantum Leap Flagship Program (MEXT Q-LEAP) Grant Number JP-MXS0118068681.
O.D.M acknowledges support by the priority program QUTIF (SPP1840 SOLSTICE) and the Cluster of Excellence `Advanced Imaging of Matter' (EXC 2056 - project ID 390715994) of Deutsche Forschungsgemeinschaft.
P.F.L. acknowledges support by National Key Research and Development Program (2017YFE0116600) and National Natural Science Foundation of China(91950202).
\\

\noindent \textbf{Author Contributions} B.X., Y.T., and E.J.T designed the experimental setup and performed the experiments. 
Y.F. contributed to the development of the CEP-stabilized Ti:sapphire laser system.
H.Y and P.F.L. performed the theoretical calculations of HHG.
B.X. and E.J.T. discussed all the experimental results and wrote the manuscript, which was polished by all authors.
E.J.T. conceived the experimental idea and supervised this project as a whole.\\

\noindent \textbf{Competing Interests} The authors declare that they have no competing financial interests.\\

\noindent \textbf{Data and materials availability:} Additional data and materials are available online.




\newpage

\noindent{\large\bf Supplementary information}\\

\begin{center}
\noindent{\Large Fully stabilized multi-TW optical waveform synthesizer for gigawatt soft-x-ray isolated attosecond pulses}\\
\vspace{\baselineskip}
\noindent{Bing Xue, Yuuki Tamaru, Yuxi Fu, Hua Yuan, Pengfei Lan, Oliver D. M\"ucke,\\Akira Suda, Katsumi Midorikawa, and Eiji J. Takahashi}\\
\end{center}

\vspace{\baselineskip}

%


\begin{center}
\includegraphics[width=0.9\linewidth]{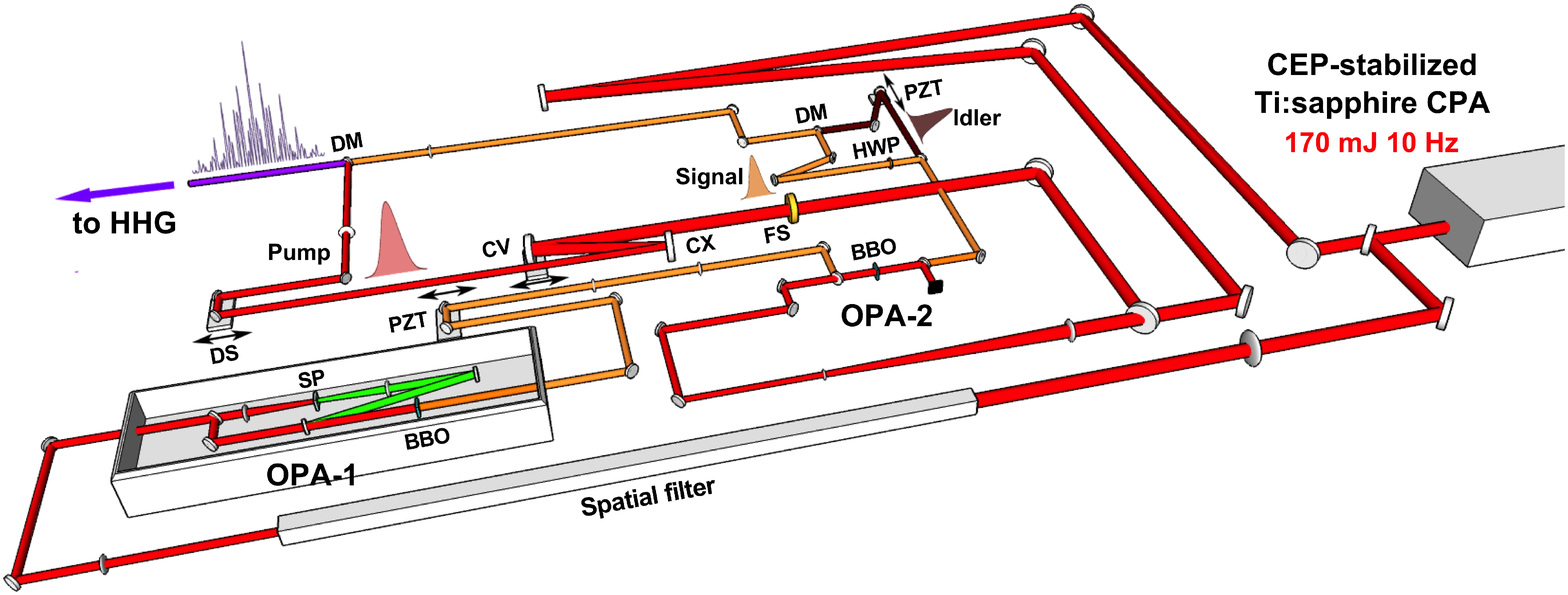}\\
\end{center}
\noindent {\bf Fig. S1. System scheme of the three-channel optical waveform synthesizer.}  CPA, chirped-pulse amplifier; CV, concave mirror; CX, convex mirror; DS, delay stage; DM, dichroic mirror; FS, fused silica plate; HWP, half-wave plate; PZT, piezo-transducer-actuated delay stage; SP, sapphire plate.\\

\begin{center}
\includegraphics[width=0.8\linewidth]{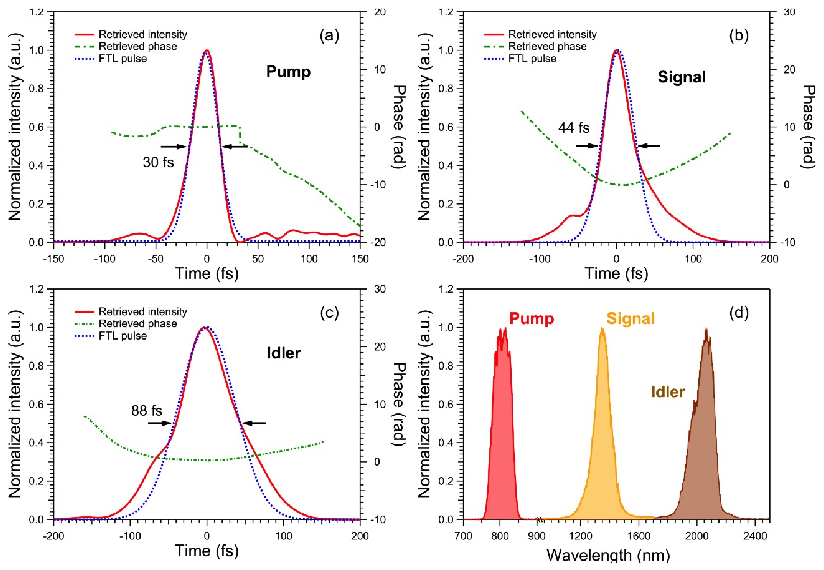}\\
\end{center}
\noindent {\bf Fig. S2. Sub-pulses composing the synthesized waveforms.} Measured durations of the pump pulse (a), signal pulse (b), and idler pulse (c). The red solid curve is the retrieved temporal intensity and the green solid curve is the retrieved temporal phase. The FTL pulse is shown as the blue dotted curve. (d) Spectra of the pump, signal, and idler sub-pulses.\\

\begin{center}
\includegraphics[width=0.8\linewidth]{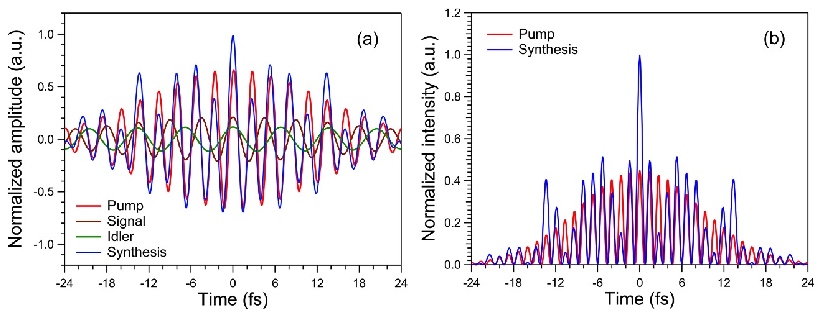}\\
\end{center}
\noindent {\bf Fig. S3.  Simulation of the synthesized electric-field waveform.} Simulated synthesized electric-field waveform $E(t)$ for the three-color case (a); and corresponding instantaneous intensity profile $E^2(t)$ (b).\\

\vspace{\baselineskip}

\centering
\noindent{\bf Table S1. Stabilized parameters of the  three-channel optical waveform synthesizer}
\begin{tabular}{cccc}
\hline
Fluctuating parameter & Pump & Signal & Idler \\
\hline
Relative delay jitter (MZ), rms & 17.3 as  & 0 & 10.8 as \\
Relative delay jitter (BOC), rms & 245 as & 0 & 137 as \\
Relative phase jitter, rms & 816 mrad & 0 & 106 mrad \\
CEP jitter, rms & 480 mrad & 572 mrad & constant \\
\hline
\end{tabular}

\end{document}